\documentclass[prl, twocolumn,english,superscriptaddress]{revtex4}
\usepackage{amsmath,amssymb,amsfonts}
\usepackage{CJK}
\usepackage{bm}
\usepackage{tikz}
\usepackage{babel}
\usepackage[colorlinks=true, letterpaper=true, pdfstartview=FitV, linkcolor=blue, citecolor=blue, urlcolor=blue]{hyperref}
\def \H {\mathcal{H}}
\def \K {\hat{\mathcal{K}}}
\def \G {\mathsf{G}}
\def \P {\mathsf{P}}

\def \i {{\mathrm{i}}}

\def \Z {\mathbb{Z}}

\def \k {\mathbf{k}}

\begin{document}

\title{Switching spinless and spinful topological phases with projective \emph{PT} symmetry}

\author{Y. X. Zhao}
\email[]{zhaoyx@nju.edu.cn}
\affiliation{National Laboratory of Solid State Microstructures and Department of Physics, Nanjing University, Nanjing 210093, China}
\affiliation{Collaborative Innovation Center of Advanced Microstructures, Nanjing University, Nanjing 210093, China}

\author{Cong Chen}
\affiliation{School of Physics, and Key Laboratory of Micro-nano Measurement-Manipulation and Physics, Beihang University, Beijing 100191, China}

\author{Xian-Lei Sheng}
\affiliation{School of Physics, and Key Laboratory of Micro-nano Measurement-Manipulation and Physics, Beihang University, Beijing 100191, China}

\author{Shengyuan A. Yang}
\address{Research Laboratory for Quantum Materials, Singapore University of Technology and Design, Singapore 487372, Singapore}

\begin{abstract}
A fundamental dichotomous classification for all physical systems is according to whether they are spinless or spinful. This is especially crucial for the study of symmetry-protected topological phases, as the two classes have distinct symmetry algebra. As a prominent example, the spacetime inversion symmetry $PT$ satisfies $(PT)^2=\pm 1$ for spinless/spinful systems, and each class features unique topological phases. Here, we reveal a possibility to switch the two fundamental classes via $\Z_2$ projective representations. For $PT$ symmetry, this occurs when $P$ inverses the gauge transformation needed to recover the original $\Z_2$ gauge connections under $P$. As a result, we can achieve topological phases originally unique for spinful systems in a spinless system, and vice versa. We explicitly demonstrate the claimed mechanism with several concrete models, such as Kramers degenerate bands and Kramers Majorana boundary modes in spinless systems, and real topological phases in spinful systems. Possible experimental realization of these models is discussed. Our work breaks a fundamental limitation on topological phases and opens an unprecedented possibility to realize intriguing topological phases in previously impossible systems.

\end{abstract}
\maketitle

Symmetry-protected topological phases have constituted one of the most active fields over the last decade and a half~\cite{Volovik2003,Hasan2010,Qi2011,Shen2012,Bansil2016,Armitage2018}. Based on mathematical tools such as the $K$ and $KO$ theories~\cite{ATIYAH1966,Karoubi1978}, rich topological phases have been proposed and classified by considering various internal and space group symmetries~\cite{Hofmmodeheckrlseriava2005,Schnyder2008,Kitaev2009,Zhao2013,Zhao2014,Chiu2016,Zhao2016,Slager_prx}.

In this endeavor, a fundamental dichotomy is to distinguish systems based on whether they are spinful or spinless. For electronic systems, this corresponds to whether spin-orbit coupling (SOC) is included or not. The two categories exhibit distinct topological classifications. The reason is that for spinful systems, due to SOC, symmetry transformations must simultaneously act on both the orbital and the spin degrees of freedom, leading to symmetry algebra distinct from spinless systems.

A prominent example is the spacetime inversion symmetry $PT$. For spinful systems, $(PT)^2=-1$, which dictates a Kramers double degeneracy at every $k$-point in the Brillouin zone (BZ). In contrast, for spinless systems, $(PT)^2=1$, which instead guarantees a {real} band structure, because one can always choose a representation with $\hat{P}\hat{T}=\hat{\mathcal{K}}$, with $\hat{\mathcal{K}}$ the complex conjugation. Each class hosts its own unique collection of topological phases~\cite{Zhao2016}.
For instance, $PT$-invariant spinful systems can realize 3D Dirac semimetals, 1D topological insulators/superconductors in class DIII; whereas spinless systems harbor real Dirac semimetals~\cite{Zhao2017}, $\Z_2$-charged nodal surfaces~\cite{Nodal_Surfaces}, nodal-line linking structures~\cite{Bzdufmmodeheckslsesiek2017,Ahn2018,Wang2019}, boundary phase transitions~\cite{Wang2020}, and etc~\cite{Ahn2019,Sheng2019,Wu2019}. A list of topological classification when including $PT$ and sublattice symmetry $S$ is presented in Table~\ref{tab:Cls_Table}.


The spin class therefore imposes a fundamental constraint on the possible topological phases that a system can realize. \emph{It is possible to break this limitation?} Namely, is it possible to realize spinful (spinless) topological phases in spinless (spinful) systems?

In this Letter, we discover an approach to achieve this possibility. The essence of our proposal is that in the presence of gauge degrees of freedom, symmetries of a system will be \emph{projectively} represented, which may completely change the fundamental algebraic structure of the symmetry group~\cite{Moore2020}.
Particularly, we show that this can be achieved by coupling to a  $\Z_2$ gauge field. Here, $\Z_2=\{\pm 1\}$ is the subgroup of the electromagnetic gauge group $U(1)$, and physically just corresponds to switching the \emph{sign} of certain hopping amplitudes.   Remarkably, we find that the projectively represented symmetry $\P T$ may satisfy $(\P T)^2=-1$ [$(\P T)^2=1$] for spinless (spinful) systems, namely, that we can switch the fundamental symmetry algebras between spinless and spinful systems. In a sense, we hence effectively make a spinful system behave as a spinless one, and vice versa.
We explicitly demonstrate our idea via several concrete models, such as Kramers degenerate
bands and Kramers Majorana boundary modes in spinless systems, and real Stiefel-Whitney topological phases
in spinful systems. Experimental realizations of these models are discussed.



Our work opens up an unprecedented possibility to switch the fundamental categories of topological systems and to achieve intriguing topological phases in previously impossible systems.

\begin{table}
	\begin{tabular}{c|cc|ccc}
		& $(PT)^2$ & $S$ & $d=1$ & $d=2$ & $d=3$\\
		\hline
		\hline
		AI  & $+$ & $0$ & $\Z_2$ & $\Z_2$ & $0$ \\
		BDI & $+$ & $[S,PT]=0$ & $\Z_2$ & $0$ & $2\Z$ \\
		CI  & $+$ & $\{S,PT\}=0$ & $\Z$ & $\Z_2$ & $\Z_2$ \\
		\hline
		AII & $-$ & $0$ & $0$ & $0$ & $0$\\
		CII & $-$ & $[S,PT]=0$ & $0$ & $0$ & $\Z$\\
		DIII & $-$ & $\{S,PT\}=0$ & $2\Z$ & $0$ & $0$
	\end{tabular}
	\caption{Topological classification table for spacetime inversion and sublattice symmetries. \label{tab:Cls_Table}}
\end{table}

{\color{blue}\textit{Projective $\P T$ symmetry}.} Let's start with a general discussion of the $PT$ symmetry.
Ordinarily, for a system consisting of particles with spin-$s$, the time-reversal symmetry satisfies $T^2=(-1)^{2s}$, and the space inversion symmetry satisfies $P^2=1$. They commute with each other, $[P,T]=0$, and therefore
\begin{equation}
(PT)^2=(-1)^{2s}.
\end{equation}

For instance, in the internal space of an electron (which is spinful), we have $(PT)^2=-1$. The common textbook explanation is that $T$ is represented by $\hat{T}=-\i\sigma_2\K$ with $T^2=-1$, while $P$ is represented by $\hat{P}=\sigma_0$ which preserves the spin. Here, $\sigma$'s are the Pauli matrices for spin. On the other hand, for spinless particles, $\hat{T}=\K$ and $\hat{P}=1$, and therefore $(PT)^2=1$.

However, in the presence of certain gauge degree of freedom, the relation $(PT)^2=(-1)^{2s}$ will be projectively represented, because the inversion is a spatial symmetry and may involve additional gauge transformations. Here, we request that the gauge flux configuration is invariant under $P$, i.e., $P$ is still a symmetry of the system. Nevertheless, the chosen gauge connections do not necessarily preserve $P$.  Then, to recover the gauge configuration, a gauge transformation $\G$ must be incorporated into the inversion. Thus, the \emph{proper} inversion actually becomes a combined operation,
\begin{equation}
\P=\G P.
\end{equation}

Specifically, for a $\Z_2$ gauge theory, $\G$ preserves $T$ and $\G^2=1$. In addition, if $P$ reverses the gauge transformations, i.e., $\G$ anti-commutes with $P$, then we have the following relations:
\begin{equation}\label{Gauge-commutations}
[\G,T]=0,\quad \{\G, P\}=0,\quad \G^2=1.
\end{equation}
It follows that $\P^2=(\G P)^2=-1$. Thus, the \emph{proper} spacetime inversion symmetry $\P T$ will satisfy a distinct algebra:
\begin{equation}\label{Projective-PT}
(\P T)^2=\P^2T^2=(-1)^{2s+1}.
\end{equation}

This is remarkable, because it shows that, with the help of $\Z_2$ gauge fields, the fundamental symmetry algebra can be exchanged for spinless and spinful systems. Consequently, their topological classifications are also exchanged. For instance, in the tenfold classification of Hamiltonian spaces, the topological phases of classes AI, BDI and CI (AII, CII and DIII) as shown in Table~\ref{tab:Cls_Table} can now be realized also by spinful (spinless) systems. Here, the tenfold classification involves the sublattice symmetry, but clearly, the discussion can be extended to other symmetries such as various crystalline symmetries as well.

Below, we present three concrete models to demonstrate our idea. More examples can be found in the Supplemental Material (SM)~\cite{Supp}.


\begin{figure}
	\includegraphics[scale=1.0]{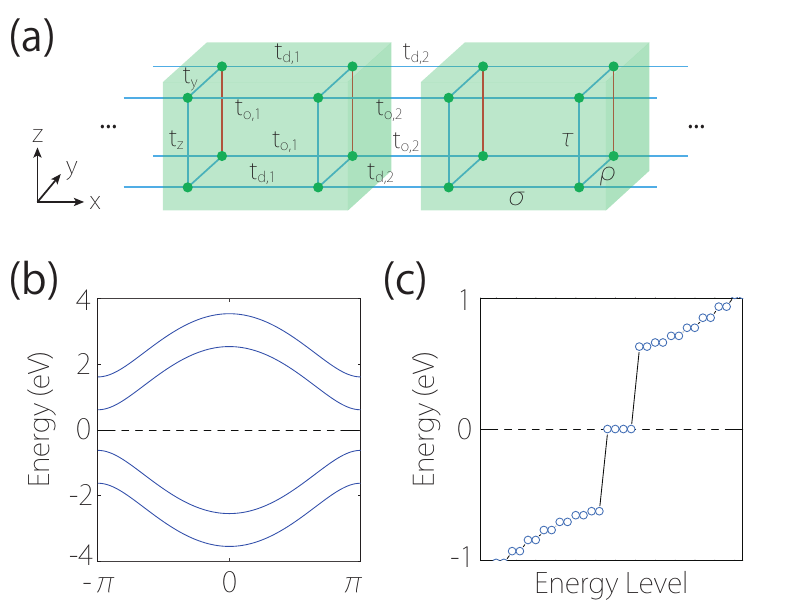}
	\caption{(a) Schematic figure of the 1D spinless chain. Each unit cell (indicated by the green cubes) contain eight sites. The hopping amplitudes along $x$ are marked in the figure. The red bonds have a negative hopping amplitude $-t_z$. (b) Calculated bulk band structure. Each band is twofold degenerate. (c) Spectrum of a finite chain with a length of 30 unit cells. The four zero-modes form two Majorana Kramers pairs. Each pair is localized at one end of the chain.  \label{1Dmodel}}
\end{figure}

{\color{blue}\textit{Kramers Majorana modes in a 1D spinless chain}.} Our first example is a 1D spinless model which realizes a class DIII topological gapped phase originally unique for spinful systems (see Table~\ref{tab:Cls_Table}).

The model is illustrated in Fig.~\ref{1Dmodel}(a). The unit cell (indicated by the shaded cube) contains eight sites, indexed by three qubits, $\rho$, $\tau$, and $\sigma$.
The inversion operator flips all three qubits, therefore in momentum space
is given by
\begin{equation}
	\hat{P}=\Gamma_{111}\hat{I},
\end{equation}
where we define
\begin{equation}
  \Gamma_{\mu\nu\lambda}=\rho_\mu\otimes \tau_\nu\otimes\sigma_\lambda,
\end{equation}
with $\mu,\nu,\lambda=0,1,2,3$, and $\hat{I}$ is the momentum inversion operator.

The $\Z_2$ gauge field is specified that each plaquette normal to the $x$ direction has a $\pi$-flux, and all others have a zero flux. One gauge-connection configuration is shown in Fig.~\ref{1Dmodel}(a), where only each red-colored bond has negative sign in the hopping amplitude.

Clearly, the flux configuration respects the $P$ symmetry, however, the gauge-connection configuration in Fig.~\ref{1Dmodel}(a) does not. To restore the original gauge configuration,
the proper inversion $\P$ should include the following gauge transformation:
\begin{equation}
	\hat{\mathsf{G}}=\Gamma_{003},
\end{equation}
which imposes a minus sign for sites on the bottom layer, and therefore, flips the sign for hopping amplitudes along $z$.
Hence, the proper inversion operator is $\hat{\mathsf{P}}=\hat{\mathsf{G}}\hat{P}=i\Gamma_{112}\hat{I}$,
and the spacetime inversion is represented by
\begin{equation}
	\hat{\mathsf{P}}\hat{T}=i\Gamma_{112}\K.
\end{equation}
Importantly, note that $P$ inverses the gauge transformation $\mathsf{G}$, such that $\{\hat{\mathsf{G}},\hat{P}\}=0$.
Therefore, according to our analysis above [see Eqs.~\eqref{Gauge-commutations} and \eqref{Projective-PT}], the projective $\mathsf{P}T$ symmetry of this spinless chain follows a modified algebra $(\mathsf{P}T)^2=-1$, a character intrinsic to spinful systems.

With the hopping amplitudes shown in Fig.~\ref{1Dmodel}(a), the tight-binding model can be written as
\begin{equation}
		\H(k)= t_y \Gamma_{010}+t_z \Gamma_{301}+\sum_{s=d,o} \begin{bmatrix}
			0 & u_s(k)\\
			u^*_s(k) & 0
		\end{bmatrix}\otimes M_s.
\end{equation}
Here, $k$ is the momentum along $x$, $M_d=\mathrm{diag}(1,0,0,1)$, $M_o=\mathrm{diag}(0,1,1,0)$, and $u_{s}(k)=t_{s,1}+t_{s,2} e^{-ik}$. 

The sublattice symmetry is represented as
$\hat{S}=\Gamma_{333}$. It can be transformed into the standard form $\Gamma_{300}$ by the unitary transformation $	U= \exp(\frac{i\pi}{4}\Gamma_{100})\exp(-\frac{i\pi}{4}\Gamma_{133})$. Accordingly, the Hamiltonian can be converted into the standard block off-diagonal form as for class DIII systems
\begin{equation}
  U\H U^\dagger=\begin{bmatrix}
			0 & Q(k)\\
			Q^\dagger (k) & 0
		\end{bmatrix}.
\end{equation}
For gapped phases in class DIII, the 1D topological invariant is given by the winding number
\begin{equation}
	\nu=\frac{1}{2\pi i}\oint dk~ \mathrm{tr} Q^{-1}(k)\partial_k Q(k).
\end{equation}
This winding number is valued in even integers $2\Z$ due to the algebraic relations $(\hat{\mathsf{P}}\hat{T})^2=-1$ and $\{\hat{\mathsf{P}}\hat{T},\hat{S}\}=0$, as proved in the SM~\cite{Supp}.

For instance, the system is nontrivial with $\nu=2$, when we set $t_y=t_z=0.5$, $t_{d,1}=t_{o,2}=2$, and $t_{d,2}=t_{o,1}=1$. The corresponding band bulk structure is shown in Fig.~\ref{1Dmodel}(b), showing a gapped spectrum. Note that although the system is spinless, each band here has a Kramers double degeneracy due to $(\P T)^2=-1$. The invariant $\nu=2$ dictates that for a $PT$-symmetric chain with an open boundary condition, there must exist a Kramers pair of Majorana modes at each boundary, which is confirmed by our result in Fig.~\ref{1Dmodel}(c). Previously, such Kramers Majorana pair is only possible for spinful systems, such as the 1D $T$-invariant $p$-wave topological superconductor. Here, we demonstrate that it can be successfully extended to spinless systems via our proposed mechanism. Since only the nearest neighbor hopping is needed here, the topological phase could be easily realized by various artificial systems.


\begin{figure}
	\includegraphics[scale=1]{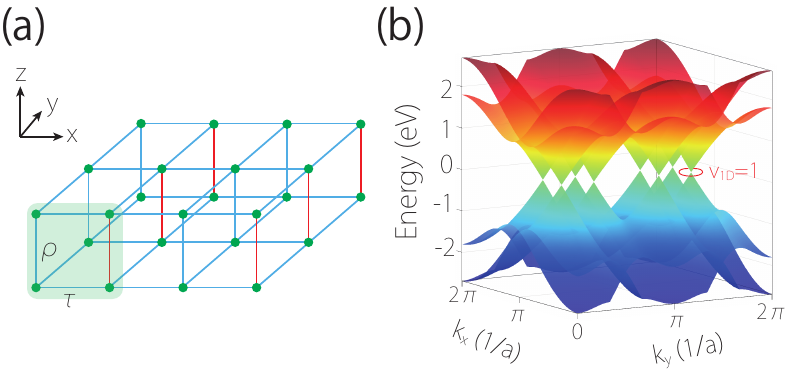}
	\caption{(a) Schematic figure for the 2D model, consisting of two layers of square lattices. The unit cell consists of four sites as marked by the green square. The red colored bonds have a negative hopping amplitude $-t_z$. (b) Bulk band structure of the model (\ref{2D_Dirac}). There are eight twofold real Dirac points at zero energy, protected by $\nu_\text{1D}=1$. Here, we take parameters $t_x=t_y=t_z=1$ and $\lambda=0.5$. \label{2Dmodel}}
\end{figure}


{\color{blue}\textit{2D real Dirac semimetal in a spinful lattice}.} In the second example, we achieve a 2D phase with real twofold Dirac points.
The real Dirac point is previously unique for spinless systems.  A famous example is graphene. It is protected by the first Stiefel-Whitney number $\nu_\text{1D}$, which just corresponds to the quantized Berry phase along a circle. Here, we will realize it in a spinful system.


As illustrated in Fig.~\ref{2Dmodel}(a), the 2D model consists of two layers of square lattices. Each square plaquette normal to the $y$ direction has a $\pi$-flux. A possible gauge configuration is shown in Fig.~\ref{2Dmodel}(a), where the red-color bonds have a negative hopping amplitude. A unit cell contains four cites, which are labeled by two qubits $\rho$ and $\tau$ [Fig.~\ref{2Dmodel}(a)], and the spin basis on each site is denoted by $\sigma$.

Clearly, the inversion
$\hat{P}=\Gamma_{110}\hat{I}$ reverses the vertical hopping amplitudes, therefore the proper inversion must include a gauge transformation
$\hat{\mathsf{G}}=\Gamma_{300}$, which anti-commutes with $\hat{P}$.
Hence, $\hat{\mathsf{P}}=i\Gamma_{210}\hat{I}$. With the standard $\hat{T}=-i\Gamma_{002}\hat{I}\K$ for a spinful system, we obtain
\begin{equation}\label{Spinful-PT}
	\hat{\mathsf{P}}\hat{T}=\Gamma_{212}\K,
\end{equation}
which satisfies the identity $(\hat{\mathsf{P}}\hat{T})^2=1$, so we have made the spinful system behave effectively as a spinless one.

With the hopping amplitudes specified in Fig.~\ref{2Dmodel}(a), our spinful lattice model is given by
\begin{multline}\label{2D_Dirac}
		\H(\bm k)=(t_x+t_x\cos k_x) \Gamma_{010}+ t_x \sin k_x \Gamma_{020}\\+2t_y \cos k_y \Gamma_{031}
		+t_z \Gamma_{130}+\lambda \Gamma_{303}.
\end{multline}
Here, the third term is an explicit SOC term. Figure~\ref{2Dmodel}(b) shows a typical band structure for this model. The spectrum contains eight twofold Dirac points at zero energy. It is easy to verify that these are real Dirac points, each protected by a $\pi$ Berry phase quantized by the projective $\P T$ symmetry.


This 2D real Dirac semimetal can be readily extended into a 3D real nodal-line semimetal by stacking its copies along the $z$ direction, e.g., by adding the following vertical hopping terms to Eq.~(\ref{2D_Dirac}):
\begin{equation}
		\H_z(k_z)= (t_z+t_z\cos k_z)\Gamma_{130}+t_z\sin k_z \Gamma_{230}.
\end{equation}
They will generate four real nodal loops in the 3D BZ~\cite{Supp}, and each loop is protected by a quantized $\pi$ Berry phase.



{\color{blue}\textit{Real Dirac point and doubly-charged loop in generalized 3D Kane-Mele model}.} The third example is a real topological phase characterized by the second Stiefel-Whitney number $\nu_\text{2D}$, corresponding to class AI in Table~\ref{tab:Cls_Table} but realized in a spinful system.

\begin{figure}
	\includegraphics[scale=1]{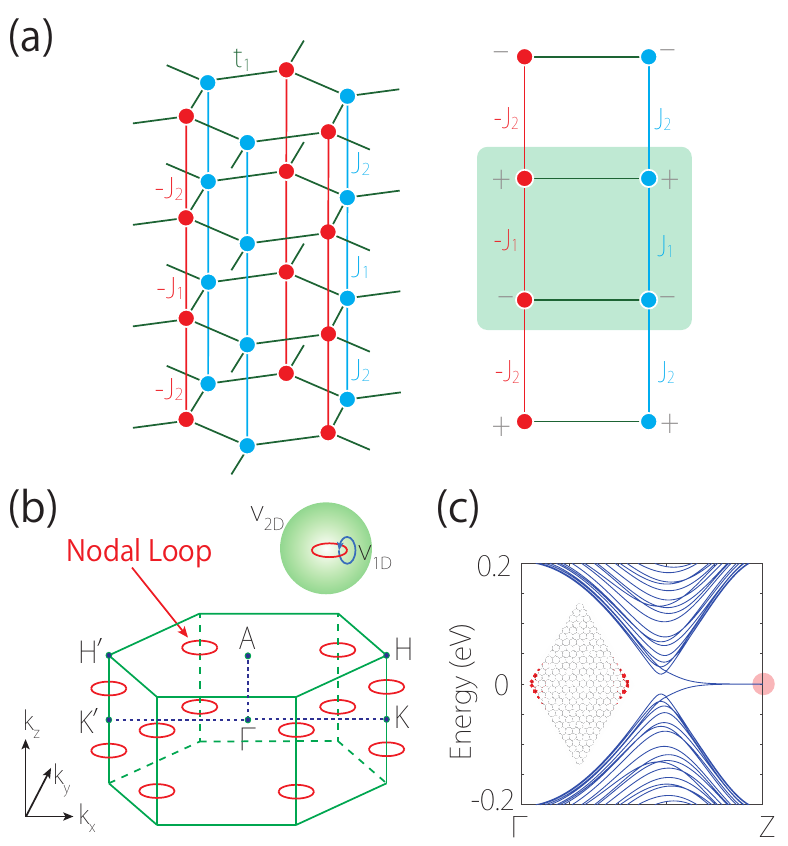}
	\caption{(a) Schematic figure for the 3D generalized Kane-Mele model. Each 2D honeycomb layer is a 2D Kane-Mele model. The red colored bonds have negative hopping amplitudes $-J_1$ or $-J_2$. The unit cell contains four sites, as indicated by the shaded square in the right panel.
(b) The bulk band structure contains four real nodal loops. Each loop is characterized by a twofold topological charge $(\nu_\text{1D}, \nu_\text{2D})=(1,1)$, as indicated by the inset. (c) Calculated spectrum of the system with a tube-like
geometry extended along $z$ and with a diamond-shaped
cross section (see inset). The hinge Fermi arc states can be clearly visualized. Here, we take parameters $t_1=1$, $t_2=0.08$, $J_1=0.3$, $J_2=0.48$, $m_1=0.4$ and $m_2=0.3$. \label{3DKM}}
\end{figure}

Our model is constructed by stacking the renowned 2D Kane-Mele model with interlayer $\pi$ fluxes. As shown in Fig.~\ref{3DKM}(a), the stacking forms a 3D graphite lattice, and the $\pi$ flux only exists for each vertical rectangular plaquette. Figure \ref{3DKM}(a) shows a possible gauge connection configuration, where again we use the red color to indicate the bonds with a negative hopping amplitude. We define a unit cell with four sites, as marked by the shaded region. These four sites are indexed by the quibits $\rho$ and $\tau$, and again the real spin is denoted by $\sigma$. Following similar analysis above, we find that the proper inversion operator $\hat\P$ must include $\hat{P}=\Gamma_{110}\hat I$ and the gauge transformation $\hat G=\Gamma_{300}$. Combined with $\hat T=-i\Gamma_{002}\hat{I}\hat{\mathcal{K}}$ for spinful systems, we find
$\hat \P \hat T=\Gamma_{212}\hat{\mathcal{K}}$, same as \eqref{Spinful-PT}.
Clearly, $(\hat \P \hat T)^2=1$, hence a spinful system is effectively turned into a spinless one.

The lattice model is given by
\begin{equation}\label{Graphite-Model}
\begin{split}
\H(\bm k)=& \chi_1(\bm k) \Gamma_{010}+\chi_2(\bm k) \Gamma_{020}+\eta(\bm k)\Gamma_{033}\\ &+\lambda_1(k_z)\Gamma_{130}+\lambda_2(k_z)\Gamma_{230}.
\end{split}
\end{equation}
Here, $\chi_1+i\chi_2=t_1\sum_{i=1}^3 e^{i\bm k\cdot \bm a_i}$, $\eta= -t_2\sum_{i=1}^3 \sin \bm k\cdot \bm{b}_i$, where $\bm a_i$'s are the three bond vectors for the honeycomb lattice, and $\epsilon_{ijk}\bm{b}_k=\bm{a}_i-\bm{a}_j$ are the in-plane vectors between second neighbors. The first line is just the Kane-Mele model, and the $\eta$ term is known as the intrinsic SOC term. The second line is the interlayer hopping, with $\lambda_1+i\lambda_2=J_1+J_2 e^{ik_z}$.

Let's treat $t_2$ and $\delta J=J_2-J_1$ as perturbations compared to $t_1$ and $J=(J_1+J_2)/2$. When $t_2=\delta J=0$, there are two independent eightfold Fermi points at the corners of the BZ. Turning on $t_2$ and $\delta J$, each Fermi point will split into two fourfold real Dirac points, each having a nontrivial second Stiefel-Whitney number $\nu_\text{2D}=1$.

This Dirac semimetal actually represents a critical state, in the sense that it is unstable in the presence of other
$\P T$-invariant perturbations. However, due to the nontrivial $\nu_\text{2D}$, the spectrum cannot be fully gapped. Instead, each Dirac point will evolve into a real nodal loop, protected by a twofold topological charge $(\nu_\text{1D},\nu_\text{2D})$ (see \cite{Supp} for more details, here, $\nu_\text{1D}$ is defined on a circle surrounding the loop).  For example, consider adding to model \eqref{Graphite-Model} the following $\P T$-invariant perturbations
$\Delta\H=m_1\Gamma_{301}+m_2\Gamma_{302}$. The resulting doubly charged real loops are illustrated in Fig.~\ref{3DKM}(b).

Distinct from the usual nodal-loop semimetal in spinful systems, a hallmark of such a $\P T$-invariant real nodal-loop semimetal is that it actually possesses a second-order topology, namely, it hosts protected hinge Fermi arcs. This is explicitly confirmed by our numerical calculations, as shown in Fig.~\ref{3DKM}(c).



{\color{blue}\textit{Discussion}.} This work reveals
an unprecedented possibility to break the fundamental limitation on topological phases by spin classes. We effectively
switch the spin character of a system in terms of the symmetry algebra. 
Here, we focused on the $P T$ symmetry. Clearly, the study can be extended to other symmetries and symmetry-protected topologies, which will open a new research field. The case of $CP$ symmetry is briefly discussed in the SM~\cite{Supp}.

For interacting systems, the required $\Z_2$ gauge field can appear as remaining discrete gauge symmetry after symmetry breaking~\cite{Sigrist-RMP,Discrete_Gauge_Theory,Krauss_Wilczek_Discrete_Gauge}, or as emergent field in strongly correlated systems like spin liquids~\cite{Anderson-emergent_Gauge,Wen-PSG,Kitaev2006,Xiao-Gang_RMP,Zhao_Second_SL,Lieb1994}. More importantly, it can be precisely engineered in
artificial systems, such as photonic/phononic crystals, circuit networks, and mechanical periodic systems~\cite{Zhang2018,Cooper2019,Optical_Lattice_RMP,Optical_Lattice_2,Ozawa2019rmp,MaGuancong2019nrp,Lu2014,Photonic_Crystal_quadrupole,Yang2015,Acoustic_Crystal,Imhof2018,Yu2020,Prodan_Spring,Huber2016}, which is briefly reviewed in the SM~\cite{Supp}. Particularly, we suggest that the bright-dark mechanism, i.e., the effective hopping amptitude of two sites through an intermediate high-energy site is negative, could be a universal method to engineer $\Z_2$ gauge configurations for artificial systems~\cite{Supp}.


\begin{acknowledgements}
	{\color{blue}\textit{Acknowledgements.}} The authors thank D. L. Deng for valuable discussions. This work is supported by  National Natural Science Foundation of China (Grants No. 11874201 and No. 12074024), the Fundamental Research Funds for the Central Universities (Grant No. 14380119), and the Singapore Ministry of Education AcRF Tier 2 (MOE2017-T2-2-108).
\end{acknowledgements}

\bibliographystyle{apsrev4-1}
\bibliography{PT_SOC_ref}

\clearpage
\newpage

\appendix




\begin{center}
	\textbf{
		\large{Supplemental Material for}}
	\vspace{0.2cm} 
	
	\textbf{
		\large{
			``Switching spinless and spinful topological phases by projective spacetime inversion symmetry" } 
	}
\end{center}

\vspace{-0.2cm}
\section{$\mathbb{Z}_2$ gauge fields in artificial systems}
Because of the fundamental importance of $\Z_2$ gauge fields in our theory, here we briefly review the appearance of $\Z_2$ gauge fields in  artificial systems.
\subsection{The dark-bright mechanism}
\begin{figure}
	\includegraphics[scale=0.5]{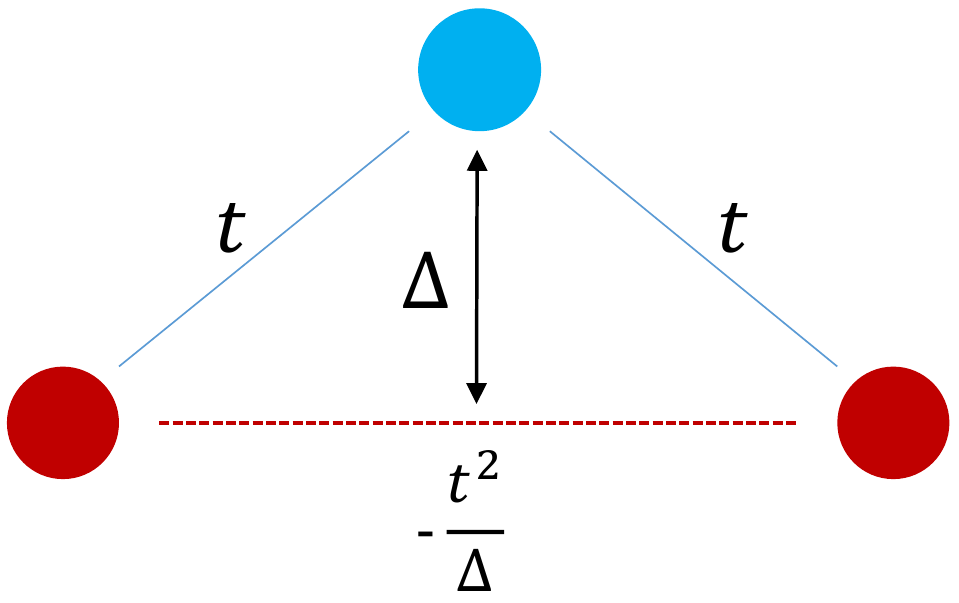}
	\caption{The effective hopping amplitude of two sites through an intermediate high-energy site.\label{fig:Dark-Bright}}
\end{figure}

The dark-bright mechanism is a simple way to engineer negative hopping amplitudes, and can be applied to any systems including electronic systems and artificial ones, such as cold atoms, photonic/acoustic crystals. Consider a three-site quantum system, illustrated in Fig.\ref{fig:Dark-Bright}. The middle site has much higher on-site energy than the other two sites $1$ and $2$. But a particle cannot hop direct from site $1$ to $2$. It has to pass through site $3$. We show in the following that there is an effective negative hopping amplitude between sites $1$ and $2$.

The Hamiltonian is given by 
\begin{equation}
	\mathcal{H}=\begin{bmatrix}
		\epsilon & 0 & t\\
		0 & \epsilon & t\\
		t & t & \epsilon+\Delta
	\end{bmatrix}
\end{equation}
where $\epsilon$ is the on-site energy for sites 1 and 2, and $\Delta$ is the energy gap above them to site $3$ with $\Delta\gg \epsilon>0$. The two low-energy levels of the system are solved as
\begin{eqnarray}
	\mathcal{E}_{D}&=&\epsilon,\quad |D\rangle=\frac{1}{\sqrt{2}}(|1\rangle-|2\rangle)\\
	\mathcal{E}_{B}&=& \epsilon-\frac{2t^2}{\Delta},\quad |B\rangle\approx \frac{1}{\sqrt{2}}(|1\rangle+|2\rangle-\frac{2t^2}{\Delta}|3\rangle)
\end{eqnarray}
The third level has much higher energy with a gap $\sim \Delta$. It is noticed that $|D\rangle$ has no component for $|3\rangle$, while $|B\rangle$ has a small component of $|3\rangle$. This justifies the their names. The low-energy effective Hamiltonian for the two low-energy levels is given by $H_{eff}=\mathcal{E}_D|D\rangle \langle D|+\mathcal{E}_B|B\rangle \langle B|$, which may also be expressed in terms of the on-site states as $H_{eff}=\sum_{ij}|i\rangle \mathcal{H}^{eff}_{ij}\langle j|$ with $i,j=1,2$. Accordingly,
\begin{equation}
	\mathcal{H}^{eff}\approx \begin{bmatrix}
		\epsilon-\frac{t^2}{\Delta} & -\frac{t^2}{\Delta} \\
		-\frac{t^2}{\Delta} & \epsilon-\frac{t^2}{\Delta}
	\end{bmatrix}.
\end{equation}
Now it is observed that there is an effective negative hopping amplitude between two low-energy sites.
\subsection{Effective gauge fields in artificial systems}
Besides the dark-bright mechanism, each artificial system has its own methods for realizing $\Z_2$ gauge fields, which have appeared in recent works for simulating topological phases.
Let us briefly review some of them noticed by us.
\begin{itemize}
	\item In photonic crystals, the sign of coupling between the site rings can be controlled by adjusting the gap between the site ring and the link-ring waveguides, so that a synthetic gauge flux threading each square plaquette can be effectively achieved~\cite{Photonic_Crystal_quadrupole}.
	\item In phononic crystals, positive and negative inter-resonator couplings are achieved by connecting the resonators with thin waveguides on different sides of each resonances nodal line~\cite{Acoustic_Crystal}.
	\item In electric-circuit arrays, negative hopping can be realized by inductors, which can be readily seen from the formula~\cite{Imhof2018,Yu2020}:
	\begin{equation}
		J_{ab}(\omega)=\i \omega (C_{ab}-\frac{1}{\omega^2 L_{ab}})
	\end{equation}
	Here $C_{ab}$ and $L_{ab}$ are capacitances and inductances, respectively.
	\item  In mechanical networks, the negative hopping can be realized by the difference of stiffness coefficients of springs~\cite{Prodan_Spring}.
\end{itemize}



\section{The Topological invariant for class DIII}
In this section, we first formulate the topological invariant for the $1$D topological insulators in class DIII in the classification table in the main text. Then, we show that the constructed model has a nontrivial topological invariant.
\subsection{Formula of the topological invariant}
The relevant symmetry operators are $PT$ and $S$, which satisfy the following algebraic relations:
\begin{equation}
	(PT)^2=-1,\quad S^2=1, \quad \{PT,S\}=0.
\end{equation}
The classifying space for Hamiltonians restricted by the symmetries is given by
\begin{equation}\label{Cls-Space}
	U/Sp,
\end{equation}
which can be understood from the following. First, we choose the representations of symmetries operators without loss of generality: $PT=i\sigma_2\K$, $S=\sigma_3$. Then, the flattened Hamiltonian is 
\begin{equation}
	\tilde{\H}=\begin{pmatrix}
		0 & \tilde{Q}\\
		\tilde{Q}^\dagger & 0
	\end{pmatrix}
\end{equation}
with
\begin{equation}
	\tilde{Q}^\dagger\tilde{Q}=1,\quad \tilde{Q}^T=-\tilde{Q}.
\end{equation}
Hence, $\tilde{Q}$ takes the decomposition form,
\begin{equation}
	\tilde{Q}=U\sigma_2 U^T,
\end{equation}
which is invariant under
\begin{equation}
	U\rightarrow USp
\end{equation}
noting that
\begin{equation}
	\sigma_2Sp\sigma_2 =Sp^*.
\end{equation}
Thus, the classifying space \eqref{Cls-Space} is justified.
The topological invariant is just the winding number
\begin{equation}
	\begin{split}
		\nu=& \frac{1}{2\pi i}\oint dk ~\mathrm{tr}\tilde{Q}^\dagger \partial_k \tilde{Q}\\
		=&\frac{1}{\pi i}\oint dk ~\mathrm{tr}U^\dagger\partial_k U \in 2\Z,
	\end{split}
\end{equation}
which is always an even integer.

\subsection{The Model}
We first summarize symmetry operators of the model. The sublattice symmetry operator for the $1$D model in Fig. 1 in the main text is 
\begin{equation}
	S=\rho_3\otimes\tau_3\otimes\sigma_3.
\end{equation}
The inversion operator is given by
\begin{equation}
	P=\rho_1\otimes\tau_1\otimes \sigma_1\hat{I},
\end{equation}
which should be followed by the gauge transformation:
\begin{equation}
	\mathsf{G}=\rho_0\otimes\tau_3\otimes \sigma_0.
\end{equation}
Thus, we obtain the actual inversion operator in momentum space under the gauge condition:
\begin{equation}
	\mathsf{P}=\mathsf{G}P=i\rho_1\otimes\tau_2\otimes \sigma_1\hat{I}.
\end{equation}
Accordingly, the spacetime inversion operator in momentum space is given by
\begin{equation}
	\mathsf{P}T=i\rho_1\otimes\tau_2\otimes \sigma_1\K.
\end{equation}
With the basis
\begin{multline}
	\psi^\dagger(k)=(a^\dagger_{+++},a^\dagger_{++-},a^\dagger_{+-+},a^\dagger_{+--},\\ a^\dagger_{-++},a^\dagger_{-+-},a^\dagger_{--+},a^\dagger_{---}),
\end{multline}
the Hamiltonian is given by
{\footnotesize
	\begin{equation}
		\begin{split}
			\H(k)= \left[\begin{matrix}
				0 & u_{d}(k) & t_z & 0 & t_y & 0 & 0 & 0\\
				u_{d}^*(k) & 0 & 0 & -t_z & 0 & t_y & 0 & 0\\
				t_z & 0 & 0 & u_{o}(k) & 0 & 0 & t_y & 0 \\
				0  & -t_z & u_{o}^*(k) & 0 & 0 & 0 & 0 & t_y \\
				t_y & 0 & 0 & 0 & 0 & u_{o}(k) & t_z & 0  \\
				0 & t_y & 0  & 0 & u_{o}^*(k) & 0 & 0 & -t_z  \\
				0 & 0 & t_y & 0 & t_z & 0 & 0 & u_{d}(k)   \\
				0 & 0  & 0 & t_y & 0  & -t_z & u_{d}^*(k) & 0 
			\end{matrix}\right]
		\end{split}	
\end{equation}}
where
\begin{equation}
	u_{d}(k)=t^1_d+t^2_d e^{ik},\quad u_{o}(k)=t^1_o+t^2_o e^{ik}.
\end{equation}
To check the symmetries, it is more appropriate to cast the Hamiltonian into the form:
\begin{equation}
	\begin{split}
		\H(k)= & t_y \Gamma_{100}+t_z \Gamma_{013}\\
		& +\frac{1}{2}(t_d^1+t_d^2\cos k)(\Gamma_{001}+\Gamma_{331})\\
		&+\frac{1}{2}t^2_d \sin k (\Gamma_{002}+\Gamma_{332})\\
		& +\frac{1}{2}(t_o^1+t_o^2\cos k)(\Gamma_{001}-\Gamma_{331})\\
		&+\frac{1}{2}t^2_o \sin k (\Gamma_{002}-\Gamma_{332}).
	\end{split}
\end{equation}
We perform the unitary transformation:
\begin{equation}
	\exp(\frac{i\pi}{4}\rho_0\otimes\tau_0\otimes\sigma_1)\exp(-\frac{i\pi}{4}\rho_3\otimes\tau_3\otimes \sigma_1),
\end{equation}
which transforms the sublattice symmetry operator into
\begin{equation}
	S\sim \rho_0\otimes\tau_0\otimes \sigma_3.
\end{equation}
Then, we have the following mappings:
\begin{equation}
	\begin{split}
		\rho_1\otimes\tau_0\otimes\sigma_0 &\mapsto \rho_2\otimes\tau_3\otimes\sigma_1\\
		\rho_0\otimes\tau_1\otimes\sigma_3 &\mapsto
		\rho_0\otimes\tau_1\otimes\sigma_2\\
		\rho_0\otimes\tau_0\otimes\sigma_1 &\mapsto
		\rho_0\otimes\tau_0\otimes\sigma_1\\
		\rho_3\otimes\tau_3\otimes\sigma_1 &\mapsto
		\rho_3\otimes\tau_3\otimes\sigma_1\\
		\rho_0\otimes\tau_0\otimes\sigma_2 &\mapsto
		\rho_3\otimes\tau_3\otimes\sigma_2\\
		\rho_3\otimes\tau_3\otimes\sigma_2 &\mapsto
		\rho_0\otimes\tau_0\otimes\sigma_2
	\end{split}
\end{equation}
With respect to $\sigma$, the Hamiltonian is block off-diagonalized into
\begin{equation}
	\H(k)\sim \begin{bmatrix}
		0 & Q(k)\\
		Q^\dagger(k) & 0
	\end{bmatrix}
\end{equation}
with
\begin{equation}
	Q(k)=\begin{bmatrix}
		t^1_d+t^2_d e^{-ik} & -it_z & -it_y & 0\\
		-it_z  & t^1_o+t^2_o e^{ik} & 0 & it_y \\
		it_y & 0 & t^1_o+t^2_o e^{ik} & -it_z\\
		0 & -it_y & -it_z & t^1_d+t^2_d e^{-ik}
	\end{bmatrix}.
\end{equation}
Then, the topological invariant is calculated as
\begin{equation}
	\nu=\frac{1}{2\pi i}\oint dk ~\mathrm{tr}Q^{-1} \partial_k Q=2,
\end{equation}
where we used the typical parameter values: $t_y=t_z=0.5$, $t_d^1=2$, $t_d^2=1$, $t_o^1=1$, $t_o^2=2$.


\section{The stacked Kane-Mele model}
\subsection{$\P T$ invariant terms}
Before constructing a specific model on the lattice, it is helpful to first count all possible $\P T$ invariant terms. For the graphite lattice here, each unit cell corresponds to an eight-dimensional Hilbert space. Accordingly, we consider the $8\times 8$ Hermitian matrices, which form a $64$D real linear space. A basis can be constructed from the tensor products of the three sets of Pauli matrices, $\rho_\mu$, $\tau_{\mu}$ and $\sigma_\mu$, with $\mu=0,1,2,3$. Hence, there are exactly 64 different tensor products, which are orthogonal under the trace inner product:
\begin{equation}
	\frac{1}{8}\mathrm{Tr}(\Gamma_{\mu\nu\lambda}\Gamma_{\mu'\nu'\lambda'})=\delta_{\mu\mu'}\delta_{\nu\nu'}\delta_{\lambda\lambda'},
\end{equation}
with
\begin{equation}
	\Gamma_{\mu\nu\lambda}=\rho_\mu\otimes\tau_\nu\otimes \sigma_\lambda.
\end{equation}
Since $\P T$ symmetry acts pointwisely (locally) in momentum space, we only need to examine which $\Gamma_{\mu\nu\lambda}$ are invariant under $\hat{\P}\hat{T}=\rho_2\otimes\tau_1\otimes \sigma_2\K$.
\begin{table}
	\begin{tabular}{cc}
		\begin{tabular}{c||c|c}
			& even & odd\\
			\hline
			~$\rho$~ & $\rho_0$ & $\rho_i$\\
			$\tau$ & $\tau_{0,1,2}$ & $\tau_3$\\
			$\sigma$ & $\sigma_0$ & $\sigma_i$
		\end{tabular} ~~&~
		\begin{tabular}{c|| c | c}
			type ($\rho\tau\sigma $) & $\P T$-invariant $\Gamma_{\mu\nu\lambda}$ & No.\\
			\hline
			eee & (0,0,0), (0,1,0), (0,2,0) & 3 \\
			eoo & $(0,3,i)$ & 3\\
			oeo & $(i,0,j)$, $(i,1,j)$, $(i,2,j)$ & 27\\
			ooe & $(i,3,0)$ & 3
		\end{tabular}
	\end{tabular}
	\caption{Left panel: Parity of the Pauli matrices under $\P T$. Right panel: List of all $\P T$-invariant tensor products. In the first column, ``e" and ``o" stand for the even and odd parity of the constituent Pauli matrices. The indices $i$ and $j$ run through 1 to 3. \label{tab:PT-invariant-terms}}
\end{table}
The parity of each Pauli matrix under $\P T$ is shown in Table~\ref{tab:PT-invariant-terms}. A $\P T$-invariant tensor product $\Gamma_{\mu\nu\lambda}$ must be a combination that consists of all even matrices, or one even plus two odd matrices. All such possibilities are summarized in Table~\ref{tab:PT-invariant-terms}, with totally 36 $\P T$-invariant terms. This table will be useful for constructing the $\P T$-symmetric models below.

\subsection{Topological characters}
To explicitly demonstrate these topological characters, let's first obtain a low-energy effective model for the critical Dirac semimetal state.
The effective model for each Dirac point should capture the four low-energy bands which are degenerate at the point. For example, the two Dirac points on $K$-$H$ each is described by
\begin{equation}\label{16}
	\mathcal{H}_D = v(q_x\gamma_1+q_y\gamma_2)+v_z q_z\gamma_3,
\end{equation}
where $\bm q$ is measured from the Dirac point, $v$ and $v_z$ are the Fermi velocities, and $\gamma_i$ are the $4\times 4$ Hermitian Dirac matrices representing the four-band basis, with $\gamma_1=\sigma_0\otimes \sigma_1$, $\gamma_2=\sigma_3\otimes\sigma_2$, $\gamma_3=\sigma_0\otimes\sigma_3$, $\gamma_4=\sigma_1\otimes\sigma_2$, and $\gamma_5=\sigma_2\otimes\sigma_2$, satisfying $\{\gamma_i,\gamma_j\}=2\delta_{ij}$. The other two points on $K'$-$H'$ are each described by $\mathcal{H}_D^*$. Since $(\P T)^2=1$, the model (\ref{16}) can always be made purely real via a unitary transformation (here by $U=e^{\gamma_2\gamma_5\pi/4}$), and the point is a real Dirac point characterized by a nontrivial Stiefel-Whitney number
\begin{equation}
	\nu_\text{2D}=\frac{1}{8\pi}\int_{S^2}\epsilon_{\mu\nu\lambda}\text{Tr}(g\mathcal{F}_R^{\mu\nu}) dn^\lambda\qquad \text{mod}\ 2,
\end{equation}
where $\mathcal{F}_R^{\mu\nu}=\partial^\mu\mathcal{A}^{\nu}-\partial^\nu\mathcal{A}^\mu+[\mathcal{A}^\mu,\mathcal{A}^\nu]$ is the real Berry curvature, $\mathcal{A}_{\alpha\beta}^\mu=\langle\alpha,\bm k|\partial^\mu|\beta,\bm k\rangle$ is the real Berry connection, with $|\alpha,\bm k\rangle$ and $|\beta,\bm k\rangle$ the real eigenstates of the two valence bands which correspond to the group $SO(2)$, $g=-i\sigma_2$ is the $SO(2)$ generator, and the integral is over a sphere $S^2$ enclosing the Dirac point.

When the critical Dirac state is driven into the nodal-line semimetal phase, e.g., by $\Delta\mathcal{H}=m_1\Gamma_{301}+m_2 \Gamma_{302}$, since $\P T$ is preserved, each nodal loop  inherits the same $\nu_\text{2D}$. In the effective model, $\Delta \mathcal{H}$ takes the form of $i\gamma_3\gamma_4 m_1+i\gamma_3\gamma_5 m_2$, one can directly verify that $\nu_\text{2D}$ is maintained. In addition, each nodal loop is also stabilized by the $\pi$ Berry phase $\nu_\text{1D}$ for any closed path encircling the loop. The discussion confirms that both the Dirac points and the nodal loops appearing in this SOC system are in fact real topological objects characterized by Stiefel-Whitney classes.

\section{Additional models}
In this section, we present additional models mentioned in the main text. We first present the $3$D hexagonal model of spinless particles that can realize a Dirac semimetal usually for spinful systems. Then, we present models on square and cubic lattices, respectively, for the $2$D Stiefel-Whitney insulator and the $3$D nodal-line semimetal, which are arguably more experimentally feasible than the model on the graphite lattice given in the main text.

\subsection{$3$D hexagonal model}
\begin{figure}
	\includegraphics[scale=1]{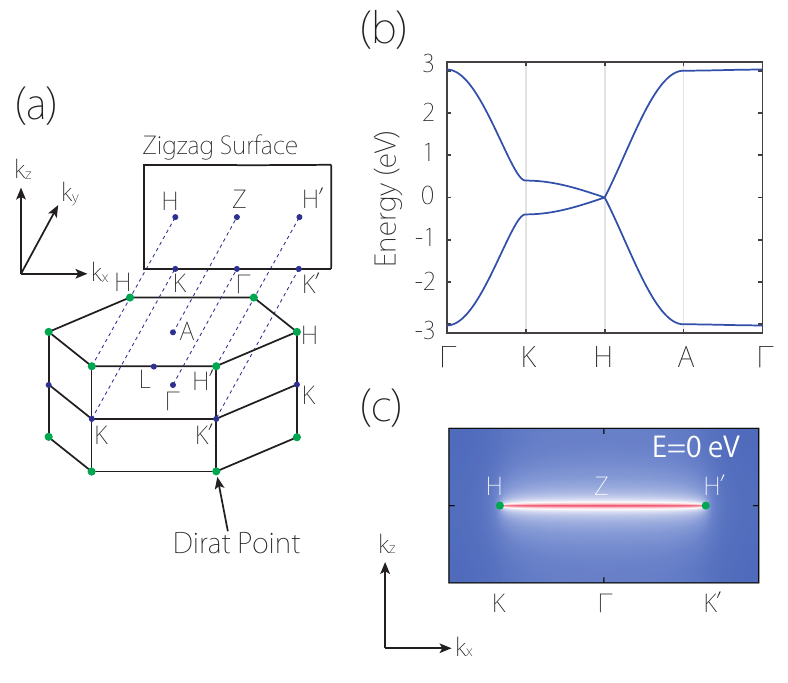}
	\caption{Spinful Dirac nodal-line semimetal by spinless particles (a) Each corner of the BZ hosts a fourfold degenerate Dirac point. (b) The bulk spectrum. Each band is twofold degenerate, and the Dirac point at $H$ has the linear dispersion relation. (c) The Fermi arc on the zigzag boundary.\label{3D_graphite}}
\end{figure}
According to common understanding, to realize genuine Dirac fermions as low-energy quasi-particles, it is necessary to have $PT$ symmetry with SOC to ensure the Kramers degenerate bands. Now, the Kramers degeneracy of each band can be protected by the projectively represented $\mathsf{P}T$ for spinless particles. Hence, we present a spinless model on the graphite lattice with $3$D genuine Dirac points.

The graphite lattice is exactly that in the main text. Each inter-layer rectangular plaquette has flux $\pi$, and the tight-binding model has only the nearest neighbor hopping. But now we consider spinless particles.  Hence, all symmetry operators have no the Pauli matrices for spins. Explicitly, the inversion operator $\hat{P}=\gamma_{11}\hat{I}$ reverses vertical positive and negative hopping amplitudes, and therefore should be accompanied by a gauge transformation $\hat{\mathsf{G}}=\gamma_{30}$. Here, $\gamma_{\mu\nu}=\rho_\mu\otimes\tau_\nu$. $\hat{\mathsf{G}}$ is inverted by $\hat{P}$, and therefore $\{\hat{\mathsf{G}},\hat{P}\}=0$. Hence, the proper inversion operator is $\hat{\mathsf{P}}=\hat{\mathsf{G}}\hat{P}=i\gamma_{21}\hat{I}$. With the standard $\hat{T}=\hat{I}\K$ for spinless particles, the proper $\mathsf{P}T$ operator is given by
\begin{equation}
	\hat{\mathsf{P}}\hat{T}=i\gamma_{21}\K,
\end{equation} 
which satisfies the identity $(\hat{\mathsf{P}}\hat{T})^2=-1$ usually for spin-$\frac{1}{2}$ particles. 
\begin{multline}\label{spinless_graphite}
	\H(\bm k)=\chi_1(\bm k) \gamma_{01}+\chi_2(\bm k) \gamma_{02}\\+\lambda_1(k_z)\gamma_{13}+\lambda_2(k_z)\gamma_{23},
\end{multline}
where all terms preserve the $\P T$ symmetry.
Here,
$\chi_1+\i\chi_2=t_1 \sum_{i=1}^3 e^{\i \bm k\cdot \bm{a}_i}$, and $\lambda_1(k_z)+\i \lambda_2(k_z)=J+Je^{\i k_z}$, where $\bm{a}_i$ with $i=1,2,3$ are the three bond vector for each hexagonal layer. 

There are two Dirac points, respectively, at the two independent corners of the $3$D 
hexagonal-prism BZ, as shown in Fig.\ref{3D_graphite}(a). The linear dispersion and twofold degeneracy can be seen from Fig.\ref{3D_graphite}(b). Each Dirac point as a crossing point of two Kramers degenerate bands is protected by the $C_3$ symmetry and the $\mathsf{P}T$ symmetry. Helical Fermi arcs are observed on the zigzag surface connecting the projections of the two Dirac points [Fig.\ref{3D_graphite}(c)].

\subsection{2D SW insulators on the square lattice}
The two-layer lattice setup and the gauge configuration are the same as those in Fig.2 in the main text. But now each particle has spin-$1/2$ described by the Pauli matrices $s_\mu$ with $\mu=0,1,2,3$.
\begin{figure}
	\includegraphics[scale=0.3]{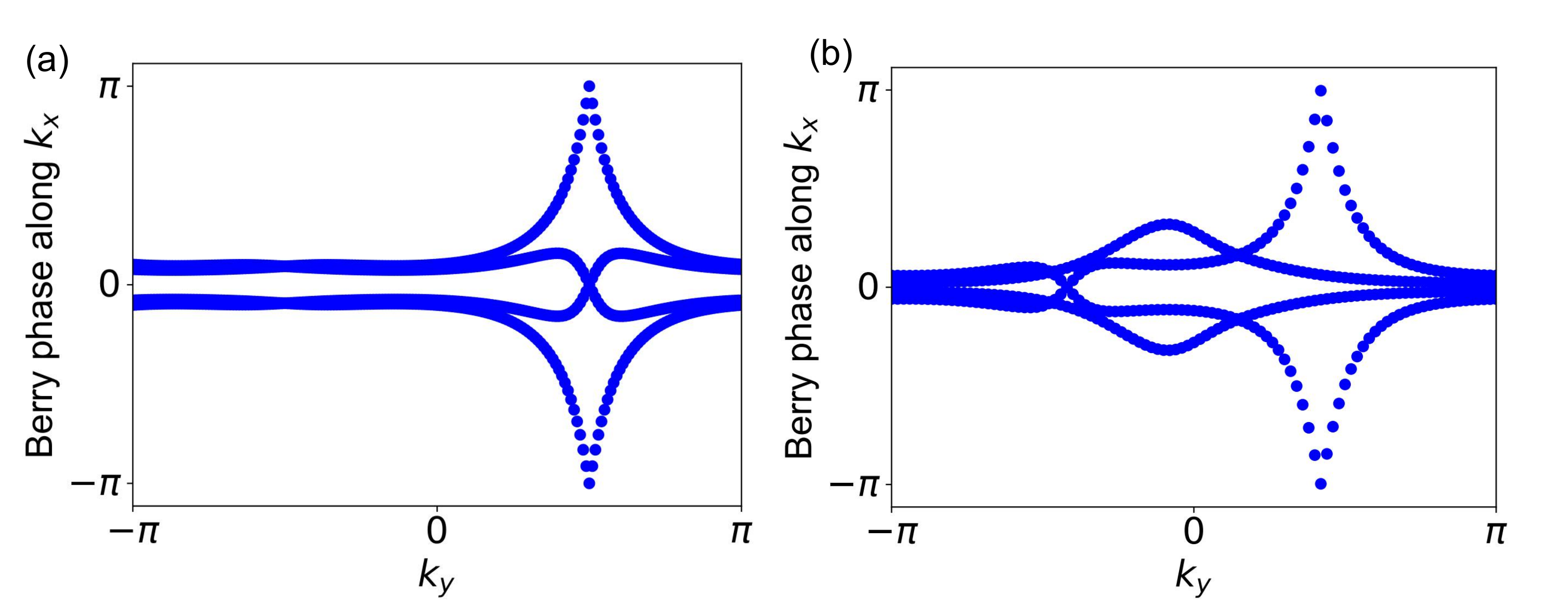}
	\caption{The Stiefel-Whitney number by the Wilson-loop spectral flow. If the number of touching points upon $\pi$ is odd (even), the topological invariant is nontrivial (trivial). Both cases correspond to the nontrivial Stiefel-Whitney number. \label{Wilson_Loops}}
\end{figure}
The first model is given by
\begin{equation}
	\begin{split}
		\H(\k)=&(t_x+t_x\cos k_x)\Gamma_{010}+ t_x \sin k_x \Gamma_{020}\\
		&+2t_y \cos k_y \Gamma_{031}+ t_z \Gamma_{102}\\
		&+(m-\sin k_x-\sin k_y) \Gamma_{033}
	\end{split}	
\end{equation}
It is topologically nontrivial for the typical parameters $t_x=t_y=1$, $t_z=0.5$, $m=1$. The Stiefel-Whitney number is computed by the Wilson loop method as shown in Fig.~\ref{Wilson_Loops} (a)~\cite{Zhao2017,Ahn2019}.
The second one is given by
\begin{equation}
	\begin{split}
		\H(\k)=&(t_x+t_x\cos k_x)\Gamma_{010}+t_x \sin k_x\Gamma_{020}\\
		&+(\lambda-2t_y \cos k_y) \Gamma_{033}+t_z \Gamma_{130}\\
		&+t_{yz} \sin k_y \Gamma_{113}
	\end{split}
\end{equation}
It has the nontrivial Stiefel-Whitney number when $t_x=t_y=t_z=1$, $\lambda=1.5$, $t_{yz}=0.5$, which can be seen from Fig.~\ref{Wilson_Loops} (b). The nontrivial physics of the $2$D Stiefel-Whitney topological insulators has been studied in Ref.~\cite{Wang2020}.

\subsection{The SW nodal-line semimetal on a cubic lattice}
\begin{figure}
	\includegraphics[scale=1]{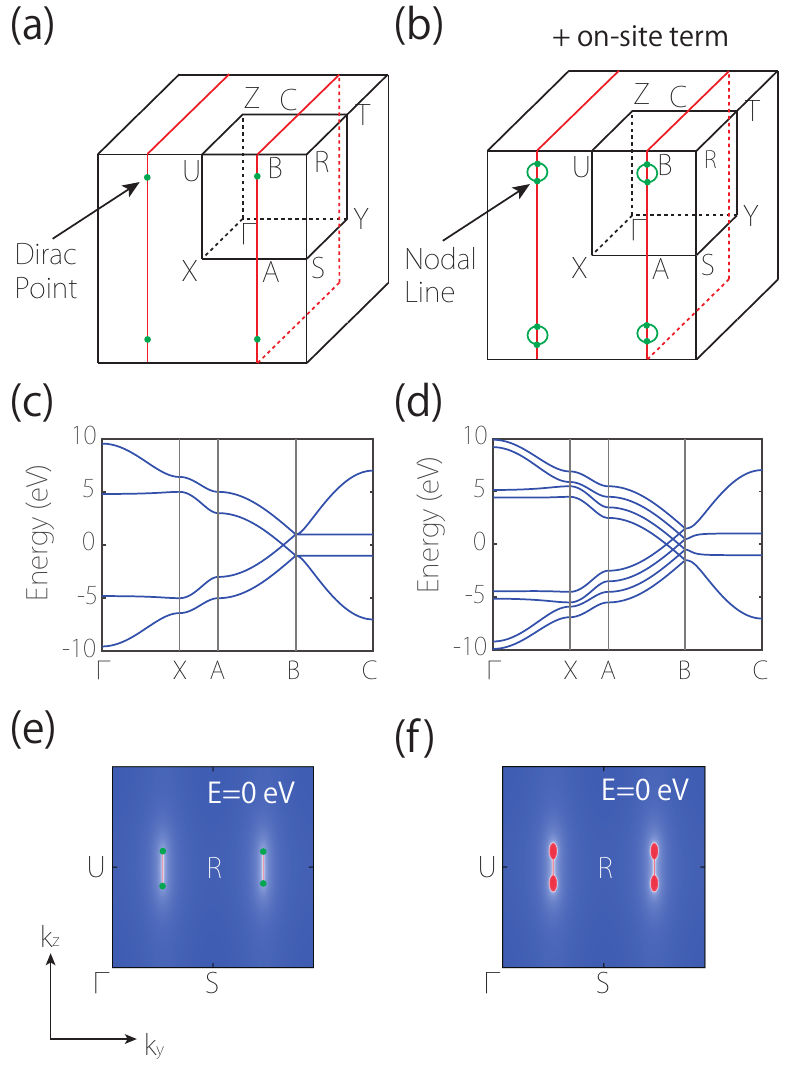}
	\caption{The numerical results for the fluxed cubic model. (a) Four real Dirac points appear in the Brillouin zone. (b) Each Dirac point is spread into a nodal line by the on-site $PT$-invariant perturbation. (c) and (d) The corresponding bulk spectra for Dirac points and nodal lines, respectively. (e) and (f) The surface states for the two cases, respectively. \label{3D_cubic_SW}}
\end{figure}
The tight-binding model on a cubic lattice, which has only the nearest hopping, and has flux-$\pi$ for all $x$, $y$, $z$-plaquettes, is given by
\begin{equation}
	\begin{split}
		\H_0(\k)=& (t_x+t_x\cos k_x)\Gamma_{010}+t_x \sin k_x \Gamma_{020}\\
		&+2t_y \cos k_y \Gamma_{330} \\ &+ (t_z+t_z\cos k_z)\Gamma_{130}+t_z \sin k_z \Gamma_{230}.
	\end{split}
\end{equation}
Here $t_x\sim t_y\sim t_z$. The spectrum contains two $8$-fold degenerate Dirac points in the Brillouin zone.

We then add the spin-orbital coupling along the $x$-direction,
\begin{equation}
	\H_1(\k)=(\lambda_{x}^1+\lambda_{x}^2\cos k_x) \Gamma_{311} +\lambda_x^2 \sin k_x\Gamma_{321},
\end{equation}
where $0<|\lambda_x^1-\lambda_x^2|<t_x$.
This term splits each eightfold degenerate Dirac point into two fourfold degenerate real Dirac points [see Fig.\ref{3D_cubic_SW}(a)]. Explicitly in Fig.\ref{3D_cubic_SW}(c), the Dirac point is a fourfold crossing point between A and B, where each band is twofold degenerate.

We further consider the on-site term
\begin{equation}
	\H_2(\k)=\alpha \Gamma_{033},
\end{equation}
where $|\alpha|<|\lambda_x^1-\lambda_x^2|$. The term spreads each fourfold real Dirac point into a nodal line normal to the $k_x$-direction [see Fig.\ref{3D_cubic_SW}(b)]. This can also be seen explicitly from the spectrum in Fig.\ref{3D_cubic_SW}(d). A typical set of parameter values are presented below.
\begin{equation}
	t_x=t_y=t_z=2,\quad \lambda_x^1=2,\quad \lambda_x^2=1,\quad \alpha=0.5.
\end{equation}
Moreover, the boundary states are worked out as well. Fig.\ref{3D_cubic_SW}(e) shows the Fermi arc states on the surface normal to the $x$-direction. And in Fig.\ref{3D_cubic_SW}(f), the projected images of the Dirac points are spread into drumhead states bounded by the projected images of nodal lines.

\section{Switching topological superconductor symmetry classes}
In the main text, we present the classification table for topological insulators considering two protecting symmetries, namely spacetime inversion symmetry and sublattice symmetry. The classifying theory in Ref.\cite{Zhao2016} can also be applied to study topological superconductors. In this case, the sublattice symmetry is replaced by the combination of time-reversal and particle-hole symmetries. Since for superconductors, the BdG Hamiltonians always have the particle-hole symmetry $C$ by construction, we further include $P$ and $T$ as protecting symmetries. Accordingly, in the framework of Ref.\cite{Zhao2016}, we particularly consider the combinations, $CP$ and $PT$, and the classifications are tabulated as Tab.\ref{tab:SC_Cls_Table}. In the presence of appropriate $\Z_2$ gauge fields, classes D, BDI and DIII will be exchanged with classes C, CII and CI, respectively. It is noteworthy that i) only topological classifications with the combined symmetries $CP$ and $PT$ are presented, while $C$ symmetry, which is always present for superconductors, may lead to additional topological classes; ii) The symbols of the tenfold symmetry classes are used to refer to the topological spaces of Hamiltonians, and should not be confused with the use for specifying superconductor classes. For instance, 1D spinless $p$-wave superconductor or the Kitaev's Majorana chain, belongs to the class-D superconductors. With $P$ symmetry considered, the $P$ operator is $\hat{P}=\tau_3 \hat{I}$ and $C$ operator is $\hat{C}=\tau_1\K\hat{I}$, with the Pauli matrices $\tau_j$ acting in the particle-hole space. Then, $\hat{P}\hat{C}=i\tau_2 \hat{K}$ with $(\hat{P}\hat{C})^2=-1$, and therefore the case corresponds to class CI in Tab.\ref{tab:SC_Cls_Table}.
\begin{table}
	\begin{tabular}{c|cc|ccc}
		& $(CP)^2$ & $(PT)^2$ & $d=1$ & $d=2$ & $d=3$\\
		\hline 
		\hline 
		D  & $+$ & $0$ & $0$ & $2\Z$ & $0$ \\
		BDI & $+$ & $+$ & $\Z_2$ & $0$ & $2\Z$ \\
		DIII  & $+$ & $-$ & $2\Z$ & $0$ & $0$ \\
		\hline 
		C & $-$ & $0$ & $0$ & $\Z$ & $\Z_2$\\
		CII & $-$ & $+$ & $0$ & $0$ & $\Z$\\
		CI & $-$ & $-$ & $\Z$ & $\Z_2$ & $\Z_2$
	\end{tabular}
	\caption{Topological classification table for superconductors with inversion and time-reversal symmetries. \label{tab:SC_Cls_Table}}
\end{table}

\end{document}